%% file: main.tex
\documentclass[aps,prb,reprint,amsfonts,longbibliography,superscriptaddress,floatfix]{revtex4-2}

\usepackage{graphicx}
\usepackage{dcolumn}
\usepackage{braket}
\usepackage{bm}
\usepackage{hyperref}
\hypersetup{colorlinks=true,linkcolor=blue,citecolor=blue,filecolor=black,urlcolor=blue}
\usepackage{subcaption}
\usepackage[normalem]{ulem}
\usepackage{amsmath}
\usepackage{amssymb}
\usepackage{physics}
\usepackage{xcolor}
\usepackage{comment}

\begin{document}

\title{Hyperuniform charge distributions and phase transitions \\ in a generalized Aubry--Andr\'e model}

\author{Lujia Xiang}
\author{Junmo Jeon}
\email{junmojeon@sophia.ac.jp}
\author{Shiro Sakai}
\email{shirosakai@sophia.ac.jp}
\affiliation{
  Physics Division, Sophia University, Chiyoda-ku, Tokyo 102-8554, Japan
}

\date{\today}

\begin{abstract}
  We show the existence of distinct inhomogeneous charge distributions and a phase transition between them in aperiodic fermion systems. Using a generalized Aubry-Andr\'e model as an example, we obtain various types of charge distributions, which we classify by means of hyperuniformity, a general mean to quantify and classify the global uniformity of spatial distributions.
  Examining various cases of filling fraction and potential strength and shape, we find that the many-body charge distribution in this model is always hyperuniform, i.e., showing an anomalously suppressed long-range fluctuation, while its hyperuniformity class depends on the localization properties of the states around the Fermi energy, irrespective of those of high-energy states.
  Namely, the change in the localization properties of these single-particle states leaves a signature in the hyperuniformity class of many-body charge distribution.
  We further show a change of the hyperuniformity class corresponds to a phase transition between inhomogeneous many-body states, and that it is of the third order.
\end{abstract}
\maketitle

\input{sections/introduction}
\input{sections/model_methods}
\input{sections/results}
\input{sections/discussion}
\input{sections/summary_and_outlook}

\begin{acknowledgments}
  This work was supported by JSPS KAKENHI Grant No.~JP25H01397, JP25H01398, JP25K24854, and JP25K24855.
\end{acknowledgments}

\appendix
\input{sections/appendix}

\bibliography{refs}
\end{document}

%% file: sections/introduction.tex
\section{Introduction}
In Landau's theory~\cite{Landau1937}, continuous phase transitions are dictated by a symmetry breaking, where an order parameter characterizes the symmetry-broken phase.
These include the ferromagnetic transition, where the time-reversal symmetry breaks, and the superconducting transition, where a gauge symmetry breaks.
In recent years, topological phase transitions have also attracted a great deal of interest as another type of phase transition, where a topological number changes~\cite{Kosterlitz-Thouless1973,Haldane1988}.
Although the symmetry- and topology-based classifications of electronic phases are useful also in aperiodic systems such as quasiperiodic and random ones, inhomogeneous electron distributions in these systems cannot be fully classified by symmetry and topology: the inhomogeneous distribution can change without changing the symmetry and topology.
For instance, the electron-charge distribution on a quasicrystalline structure~\cite{Shechtmann1984,Levine-Steinhardt1984} exhibits different spatial patterns depending on temperature, the strength of electron-electron interaction and so on~\cite{PhysRevB.105.054202,PhysRevB.105.205138}.
Natural questions then arise: Can the change of the inhomogeneous spatial pattern be the phase transition? If so, what characterizes each phase?

These questions are particularly relevant to recent experimental discoveries of various electron-ordered phases in quasicrystals, such as superconductivity~\cite{Kamiya2018,Tokumoto2024}, ferromagnetism~\cite{Tamura2021} and antiferromagnetism~\cite{Tamura2025}. Theoretical studies of superconductors and magnets in quasicrystals have indeed revealed various intriguing spatial patterns of superconducting order parameter~\cite{PhysRevB.95.024509,PhysRevB.100.014510,PhysRevResearch.1.022002,PhysRevB.102.115108,PhysRevB.104.144511,PhysRevB.112.174511} and local magnetic moment~\cite{PhysRevLett.90.177205,PhysRevB.75.212407,10.1140/epjb/e2012-21003-x,PhysRevB.96.214402,PhysRevB.102.115125,PhysRevLett.80.2717,pnas.2112202118,PhysRevMaterials.4.024417}.
The above questions on the charge distribution apply to these distributions, too.

In order to define distinct phases and discuss the phase transition between them, we need to quantify and classify the global feature of these inhomogeneous distributions. Hyperuniformity (HU)~\cite{PhysRevE.68.041113,TORQUATO20181} may offer such a framework. As will be elaborated in Sec.~\ref{subsec:Model_and_Methods_HU}, this framework looks at a spatial fluctuation of the relevant density distribution and classifies it based on its long length-scale behavior.
The charge distributions in quasiperiodic systems have indeed been shown to be hyperuniform~\cite{PhysRevB.105.054202}. In particular, in the Aubry--Andr\'e model~\cite{Aubry1980}, the HU class of the many-body charge distribution changes at the transition point from the extended to localized single-particle states. Moreover, the phase transition was found to be of the third order~\cite{PhysRevResearch.4.033241}. These results indicate that the HU framework is useful in defining distinct inhomogeneous electron phases in aperiodic systems.

Nevertheless, key challenges remain in establishing the HU framework as a general classification scheme for distinct inhomogeneous electron phases in aperiodic systems. Since all single-particle states share the same localization characteristics for a given set of Hamiltonian parameters in the Aubry--Andr\'e model, the validity of the existing HU classification becomes questionable when the charge density is composed of states with different localization characteristics arising from the presence of mobility edges. Thus, establishing a more comprehensive HU classification applicable to systems exhibiting mobility edges~\cite{PhysRevLett.114.146601,PhysRevLett.125.196604,PhysRevLett.62.2714,PhysRevB.113.L020201} is an important task.

In this paper, we address above issue by applying the HU framework to a generalized Aubry--Andr\'e (GAA) model~\cite{PhysRevLett.114.146601}.
Unlike the Aubry--Andr\'e model, the GAA model can possess a mobility edge, allowing extended and localized single-particle states to coexist and simultaneously contribute to the charge density.
We find that the charge distribution in the GAA model is always hyperuniform, while the HU class depends on whether the many-body ground state is gapped or gapless, as well as the localization properties of single-particle states around the Fermi level.
Specifically, in the gapped phases, the charge density always exhibits the same HU class. In the gapless phases, the HU class depends on the nature of the states at the Fermi energy: when the Fermi energy lies in the extended regime, the charge distribution belongs to the same HU class as in the gapped phase, whereas it exhibits a distinct HU class in the localized and critical regimes, the latter characterized by multifractal wave functions at the mobility edge.
These results naturally extend previous study of the Aubry--Andr\'e model~\cite{PhysRevResearch.4.033241} to systems exhibiting mobility edge.

Remarkably, the HU class in the gapless phases is determined solely by the localization properties of the single-particle states near the Fermi energy, regardless of those of states far from the Fermi energy.
Thus, the HU class can be regarded as a direct fingerprint of the localization character of low-energy single-particle states in gapless phases. Furthermore, when the mobility edge passes through the Fermi energy as model parameters varies, the change in the localization properties of the single-particle states from extended to localized (or vice versa) near the Fermi energy is accompanied by a change in the HU class, which results in a phase transition between inhomogeneous many-body states. Furthermore, this change in the HU class is characterized as a third-order phase transition, thereby extending the corresponding observation in the Aubry--Andr\'e model to systems with mobility edges.

The rest of the paper is organized as follows.
In Sec.~\ref{sec:Model_and_Methods}, we introduce the GAA Hamiltonian and describe the methods used to characterize single-particle localization properties, many-body charge distributions, hyperuniformity, and phase transition.
In Sec.~\ref{subsec:results_Df} and Sec.~\ref{subsec:results_HU}, we review the localization characteristics in the GAA model and study HU of the charge distribution with several respective parameter sets.
In Sec.~\ref{subsec:results_phase_transition}, we investigate the nature and order of the phase transition between different HU classes. We discuss our results in Sec~\ref{sec:discussion} and summarize this paper in Sec~\ref{sec:summary_and_outlook}.

%% file: sections/model_methods.tex
\section{Model and Methods}
\label{sec:Model_and_Methods}

\subsection{Generalized Aubry--Andr\'e model}
\label{subsec:Model_and_Methods_GAA}

We consider a tight-binding model with a local site-dependent potential $V_i$ in one dimension. The Hamiltonian reads
\begin{equation}
  \hat{H}=t\,\sum_{i}\left(\hat c_{i+1}^\dagger \hat c_i+\text{H.c.}\right)+\sum_{i}V_i\,\hat c_i^\dagger \hat c_i,
  \label{eq:gaa_H}
\end{equation}
where $\hat c_i^{\dagger}$ ($\hat c_i$) is the creation (annihilation) operator of a spinless fermion at site $i$ and the parameter $t$ denotes the nearest-neighbor hopping amplitude.
The generalized Aubry--Andr\'e model, introduced in Ref.~\cite{PhysRevLett.114.146601}, is given by the potential,
\begin{equation}
  V_i=\lambda\,\frac{\cos\left(2\pi \beta i+\phi\right)}{1-\alpha\cos\left(2\pi \beta i+\phi\right)},
  \label{eq:gaa_potential}
\end{equation}
where $\vert \alpha\vert<1$, $\beta$ is an irrational number, typically taken as the inverse golden mean $(\sqrt{5}-1)/2$, and $\phi$ is a phase shift which we set to zero throughout this paper. The parameter $\lambda$ controls the strength of the onsite potential. At $\alpha=0$, the model is reduced to the Aubry--Andr\'e--Harper model~\cite{Aubry1980,harper-1955}, which admits self-duality between real and momentum spaces, while nonzero $\alpha$ introduces higher-harmonic components into the quasiperiodic potential, breaking the self-duality.
Notably, the GAA model possesses a mobility edge $E_M$ given by~\cite{PhysRevLett.114.146601}
\begin{equation}
  \alpha E_M = 2 \, \mathrm{sgn}(\lambda) \left( |t| - |\lambda|/2 \right).
  \label{eq:mobility_edge}
\end{equation}

For numerical calculations, we consider finite-size systems of $N=F_n$ sites with periodic boundary conditions, where $F_n$ is the $n$th Fibonacci number, and replace the irrational number $\beta$ with $F_{n-1}/F_n$, which converges to the inverse golden mean as $n$ goes to infinity.
We set hopping amplitude $t = 1$ as the unit of energy and the lattice spacing $a=1$ as the unit of length. The system size is $N=F_{20}=6765$ unless otherwise mentioned.

To distinguish gapped and gapless phases in a finite-size system, we consider the energy spacing, $\Delta E$ at the Fermi energy $E_F$. Here, $\Delta E = \mathrm{inf}S_+-\mathrm{sup}S_-$, where $S_\pm=\{E\in s(\hat{H})\vert E> (<) \ E_F\}$ and $s(\hat{H})$ is the single-particle spectrum of $\hat{H}$. Note that $\Delta E$ approaches finite (zero) in thermodynamic limit for gapped (gapless) phase, respectively. Thus, we regard the system as gapped (gapless) when $\Delta E$ stabilizes (decreases) as $N$ increases.

\subsection{Inverse participation ratio and fractal dimension}
\label{subsec:Model_and_Methods_IPR_Df}

We investigate the localization characteristics of single-particle eigenstates, $\ket{\psi_{m}}$ with energy $E_m$ of $\hat{H}$, in terms of inverse participation ratio (IPR) and fractal dimension.
For eigenstate $\ket{\psi_{m}}$ with site-resolved wave function defined by $\psi_{m}(i)=\langle i\vert\psi_m\rangle$, the IPR is given by
\begin{equation}
  \mathrm{IPR}\!\left(\psi_{m}\right)=\frac{\sum_i |\psi_{m}(i)|^4}{\left[\sum_i |\psi_{m}(i)|^2\right]^2}.
  \label{eq:ipr}
\end{equation}
In our numerical calculations, when $\ket{\psi_m}$ belongs to a degenerate subspace $\mathcal{D}$, we use the average probability density $\rho_\mathcal{D}(i)=\frac{1}{|\mathcal{D}|}\sum_{p\in \mathcal{D}} |\psi_p(i)|^2$ and correspondingly
\begin{equation}
  \mathrm{IPR}({\psi_m})=\frac{\sum_i \rho_\mathcal{D}(i)^2}{\left[\sum_i \rho_\mathcal{D}(i)\right]^2}.
\end{equation}
For extended states, the IPR scales as $\mathcal{O}(1/N)$, which vanishes in the thermodynamic limit. In contrast, for localized states, the IPR is of order $\mathcal{O}(1)$ and remains finite in the thermodynamic limit.
To capture these scaling characteristics of IPR, we consider
\begin{equation}
  D_f (\psi_m,N) =-\frac{\ln \mathrm{IPR}\!\left(\psi_{m}\right)}{\ln N},
  \label{eq:D2}
\end{equation}
which gives the fractal dimension in the limit of $N\to\infty$.
Note that $D_f$ approaches unity (zero) in the thermodynamic limit for extended (localized) states.
In addition, when $D_f$ approaches an intermediate value between zero and unity in the thermodynamic limit, the state is called critical.

\subsection{Charge density distribution}
\label{subsec:Model_and_Methods_charge_density}

To investigate how localization characteristics of single-particle wave functions, characterized by the IPR and fractal dimension, manifest in many-body states, we explore the spatial distribution of the many-body charge density for different filling fractions $\nu$ and $\alpha$. Here, $\nu$ is defined as the ratio of the number of occupied single-particle states to the total number of single-particle states $N$, i.e., the average charge density of the system. At
zero temperature, the many-body charge density is obtained by summing the contributions from all occupied single-particle states.
In detail, the charge density at site $i$ is given by
\begin{equation}
  n_i \equiv \langle \hat c_i^{\dagger} \hat c_i\rangle
  = \sum_{E_{m}<E_F}
  \langle \psi_{m}|\hat c_i^{\dagger} \hat c_i|\psi_{m}\rangle,
  \label{eq:charge_distribution}
\end{equation}
where $E_F$ denotes the Fermi energy.

\subsection{Hyperuniformity}
\label{subsec:Model_and_Methods_HU}
\begin{table}
  \centering
  \caption{\label{tab:hu_class}\protect\raggedright
  HU classification in one dimension, characterized by large-$R$ behaviors of $\bar{B}(R)$ and $\mathrm{d}\bar{B}(R)/\mathrm{d}\ln R$, and small-$k$ scalings of $S(k)$ and $Z(k)$.}
  \begin{ruledtabular}
    \begin{tabular}{ccccc}
      Class & $\bar{B}(R)$ & $\mathrm{d}\bar{B}(R)/\mathrm{d}\ln R$ & $S(k)\sim k^\gamma$ & $Z(k)\sim k^\delta$ \\
      \hline
      I  & $\to \mathrm{const.}$ & $\to 0$ & $\gamma>1$ & $\delta>2$ \\
      II & $\sim \ln R$ & $\to \mathrm{const.} > 0$ & $\gamma=1$ & $\delta=2$ \\
      III & $\sim R^{1-\gamma}$ & $\to\infty$ & $0<\gamma<1$ & $1<\delta<2$
    \end{tabular}
  \end{ruledtabular}
\end{table}

The hyperuniformity (HU) framework classifies real-space physical quantities, such as charge density, based on the scaling behavior of their long-range fluctuations~\cite{PhysRevE.68.041113,TORQUATO20181,Torquato2026Weighted}. Let us consider a circular window of radius $R$ centered at $\vec{r}_c$ and the local charge density inside the window,
\begin{equation}
  N(R,\vec{r}_c)=\sum_i n_i\,\Theta\!\left(R-|\vec{r}_i-\vec{r}_c|\right),
\end{equation}
where $\vec{r}_i$ denotes the position of site $i$, and $\Theta$ is the Heaviside step function. $N(R,\vec{r}_c)$ fluctuates with $\vec{r}_c$, and the magnitude of these fluctuations generally depends on $R$. We investigate the large-$R$ scaling behavior of the variance defined by
\begin{equation}
  \sigma^2(R)=\overline{N(R)^2}-\overline{N(R)}^2,
\end{equation}
where the overline denotes an average over the window center $r_c$, sampled throughout the system. The number of samples we used is 4000.
For Poisson charge distributions in $d$-dimensions, $\sigma^2(R)$ follows the volume law, $\sigma^2(R)\sim R^d$, whereas a system with $\lim_{R\to\infty}\sigma^2(R)/R^d=0$ is called hyperuniform.

Let us focus on one-dimensional systems ($d=1$), where we can expand the variance as $\sigma^2(R)=AR+B+\mathcal{O}(R^{-1})$ for $R\gg1$. With
\begin{equation}
  A(R) \equiv \frac{\sigma^2(R)}{R},
  \label{eq:A(R)}
\end{equation}
the system is defined to be hyperuniform when $A = \lim_{R\to \infty} A(R) = 0$. In this case, $\sigma^2(R)$ approaches $B$ for large $R$. To suppress oscillations irrelevant to the large-$R$ scaling behaviors of $\sigma^2(R)$, we define $\bar B(R)$ as the average over different window sizes,
\begin{equation}
  \bar B(R) \equiv \frac{1}{R} \int_0^R \sigma^2(R') \, dR'.
  \label{eq:B(R)}
\end{equation}
The large-$R$ behavior of $\sigma^2(R)$, characterized by $\bar{B}(R)$, can be used to further classify hyperuniform systems into different classes. When $\bar{B}(R)$ converges to a constant value in the large-$R$ limit, the system is called class-I hyperuniform.
If $\bar{B}(R)$ grows logarithmically for large $R$, i.e., $\bar{B}(R)\sim \ln R$, the system is called
class-II hyperuniform. Otherwise, if $\bar{B}(R)$ grows as $\bar{B}(R)\sim R^{1-\gamma}$ with $0<\gamma<1$, the system is called class-III hyperuniform. Therefore, in the large-$R$ limit, one can distinguish HU classes by analyzing the slope of $\bar{B}(R)$ with respect to $\ln R$, namely $\mathrm{d}\bar{B}(R)/\mathrm{d} \ln R$. Specifically, this vanishes for class-I, approaches a positive constant for class-II, while diverges for class-III HU (see Table \ref{tab:hu_class}).

Fourier space provides equivalent description of HU. Let us consider the Fourier transformation of the charge-density fluctuation as $\delta n(k)=\sum_j(n_j-\nu)e^{-ikj}$. Note that the filling fraction $\nu$ is equal to the average charge density, i.e., $\nu=\frac{1}{N}\sum_j n_j$. The structure factor is defined as
\begin{equation}
  S(k) = \frac{1}{N}\left|\delta n(k)\right|^2 = \frac{1}{N}\left|\sum_{j}(n_j-\nu) e^{-ikj}\right|^2.
\end{equation}
The system is called hyperuniform when $S(k)$ vanishes as $|k|$ goes to zero, i.e., $\lim_{|k|\to 0} S(k) = 0$. Additionally, the small-$k$ scaling of the structure factor, $S(k) \sim k^\gamma$ further distinguishes different HU classes. In detail, class I (II) corresponds to $\gamma>1$ ($\gamma=1$), respectively, while $0<\gamma<1$ indicates class III (see Table~\ref{tab:hu_class}).

Since $S(k)$ consists of dense Bragg peaks in quasiperiodic systems, it is more convenient to use the integrated intensity function,
\begin{equation}
  Z(k) = 2 \int_0^{k} S(k') \, dk',
  \label{eq:Z(k)}
\end{equation}
which is smoother than $S(k)$.
From the small-$k$ dependence of $Z(k)$, we can extract the exponent $\gamma$, as the small-$k$ scaling of $S(k)\sim k^\gamma$ is equivalent to $Z(k) \sim k^{\gamma+1}$ \cite{PhysRevB.95.054119}.

In the previous study of Aubry--Andr\'e model \cite{PhysRevResearch.4.033241}, HU was introduced to characterize charge-density distributions.
However, unlike the Aubry--Andr\'e model, where all single-particle states are either extended or localized (or critical) for a given $\lambda$, the GAA model possesses a mobility edge, making it possible for occupied single-particle states to have different localization characteristics.
Hence, both localized and extended states would contribute to the charge density given by Eq.~\eqref{eq:charge_distribution}.
Thus, extending the HU classification established for the Aubry--Andr\'e model to the GAA model is crucial for achieving a comprehensive understanding of the HU classification of many-body states.

\subsection{Total energy and phase transition}
In the GAA model, the mobility edge separates localized and extended single-particle states, potentially accompanied by a change in the HU class similar to the Aubry--Andr\'e model. To characterize the associated quantum phase transition, we explore the critical behavior near the mobility edge. Specifically, at fixed filling fraction $\nu$, we investigate the criticality of the total energy per site,
\begin{equation}
  E_{\mathrm{tot}} (\lambda,\alpha) \equiv \frac{1}{N} \sum_{E_{m} < E_F} E_{m}
\end{equation}
as a function of $\lambda$ and $\alpha$, by numerically calculating its partial derivatives.

%% file: sections/results.tex
\section{Results}
\label{sec:results}
\begin{figure*}
  \begin{minipage}[b][0.455\textheight][t]{0.49\textwidth}
    \centering
    \includegraphics[width=\linewidth,trim=3pt 0 0 0,clip]{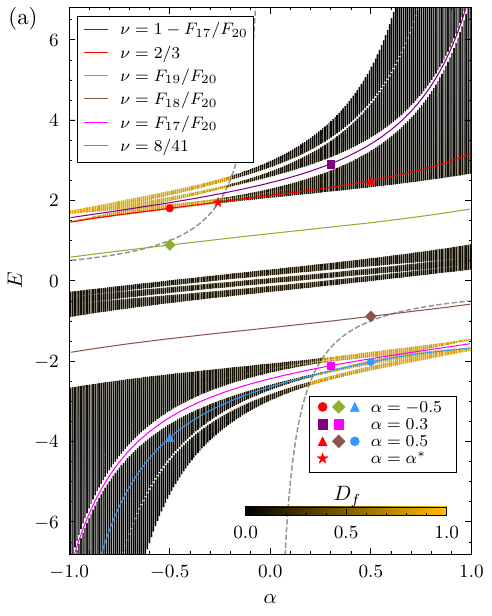}
  \end{minipage}\hfill
  \begin{minipage}[b][0.455\textheight][t]{0.49\textwidth}
    \centering
    \includegraphics[width=\linewidth,trim=0 8pt 0 0,clip]{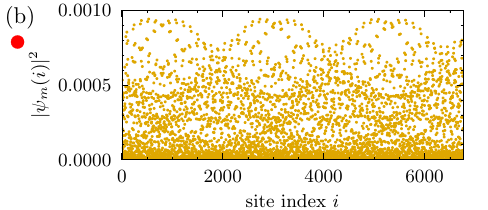}
    \\[1.5pt]
    \includegraphics[width=\linewidth,trim=0 8pt 0 0,clip]{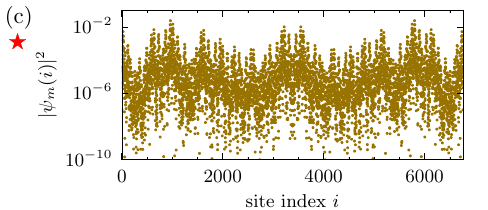}
    \\[1.5pt]
    \includegraphics[width=\linewidth]{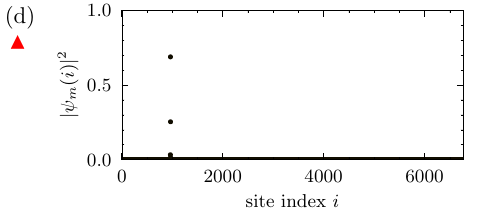}
  \end{minipage}
  \caption{\label{fig:figure_A1}\protect\raggedright Localization characteristics of single-particle eigenstates in the GAA model for $\lambda=2.5$. (a) Landscape of fractal dimension $D_f$, where gold (black) color indicates extended (localized) states.
    Gray dashed curves represent the mobility
    edges, and colored solid curves indicate the Fermi energies $E_F$ for different filling fractions $\nu$. Different symbols are used to indicate representative $(\alpha, \nu)$ points in the regions with different characteristics.
    In gapped regions, squares (diamonds) denote cases where the lowest unoccupied and highest occupied states have the same (different) localization properties. In gapless regions, circles (triangles) denote extended (localized) states at $E_F$. The red star denotes critical states at $E_F$ with $\alpha^{*}=-0.25649024$. These points are used for further analysis in the following panels and Figs.~\ref{fig:figure_B1} and \ref{fig:figure_critical}.
  (b)--(d) Real-space probability density distributions of the highest occupied state for the $(\alpha, \nu)$ points marked by the red circle, red star, and red triangle in panel (a).}
\end{figure*}

\begin{figure*}
  \begin{tabular}{cccc}
    \includegraphics[width=0.495\columnwidth]{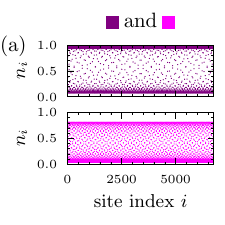} &
    \includegraphics[width=0.495\columnwidth]{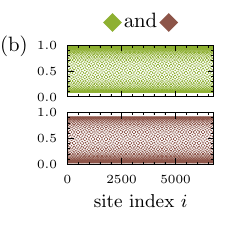} &
    \includegraphics[width=0.495\columnwidth]{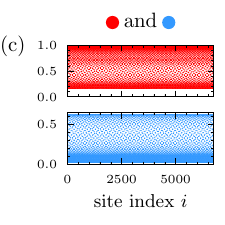} &
    \includegraphics[width=0.495\columnwidth]{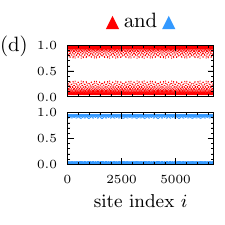} \\
    \includegraphics[width=0.495\columnwidth]{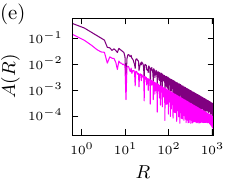} &
    \includegraphics[width=0.495\columnwidth]{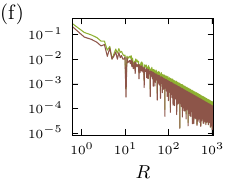} &
    \includegraphics[width=0.495\columnwidth]{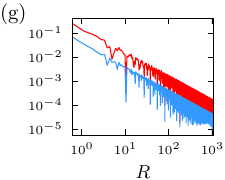} &
    \includegraphics[width=0.495\columnwidth]{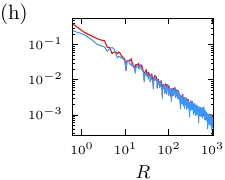} \\
    \includegraphics[width=0.495\columnwidth]{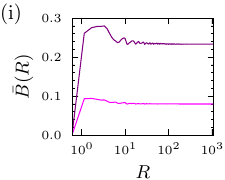} &
    \includegraphics[width=0.495\columnwidth]{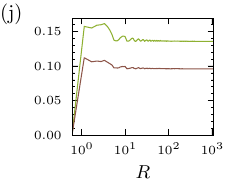} &
    \includegraphics[width=0.4955\columnwidth]{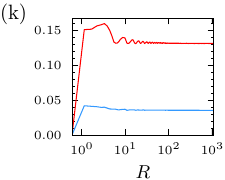} &
    \includegraphics[width=0.495\columnwidth]{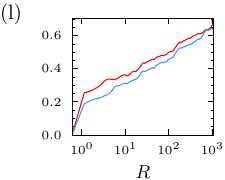}
  \end{tabular}
  \caption{\label{fig:figure_B1}\protect\raggedright Hyperuniformity classification of the GAA model for $\lambda=2.5$. For different $(\alpha, \nu)$ points indicated by different symbols in Fig.~\ref{fig:figure_A1}(a), (a)--(d) the charge-density distributions, (e)--(h) log-log plot of $A(R)$ as a function of $R$ and (i)--(l) $\bar{B}(R)$ as a function of $R$. The first and second (third and fourth) columns represent the cases of gapped (gapless) phases.
  In gapped phases, (a,e,i) represent cases where the states just above and below the Fermi energy share identical localization properties (both extended or both localized), while (b,f,j) represent mixed cases (one extended and one localized). For gapless phases, (c,g,k) represent cases where the states around the Fermi energy are extended, while (d,h,l) represent cases where those states are localized. }
\end{figure*}

\begin{figure}
  \sloppy
  \begin{tabular}{@{}cc@{}}
    \includegraphics[width=0.5\columnwidth]{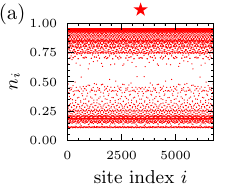} &
    \includegraphics[width=0.5\columnwidth]{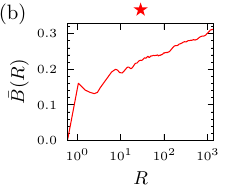} \\
  \end{tabular}
  \caption{\label{fig:figure_critical}\protect\raggedright
    (a) Charge-density distribution and (b) $\bar{B}(R)$ when the mobility edge is located between the highest occupied and lowest unoccupied states, for $\nu=2/3$, $\lambda=2.5$ and $\alpha=-0.25649024$.
  }
  \fussy
\end{figure}

\begin{figure*}
  \begin{minipage}[b][0.455\textheight][t]{0.49\textwidth}
    \centering
    \includegraphics[width=\linewidth,trim=3pt 0 0 0,clip]{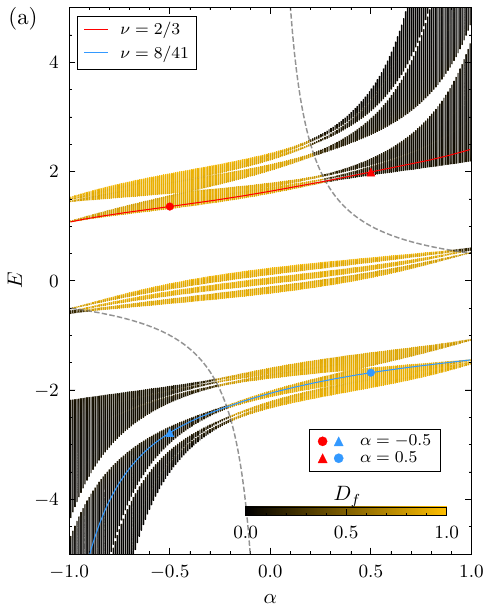}
  \end{minipage}\hfill
  \begin{minipage}[b][0.45\textheight][t]{0.49\textwidth}
    \centering
    \vspace*{-14pt}
    \begin{tabular}{@{}cc@{}}
      \includegraphics[width=0.49\linewidth]{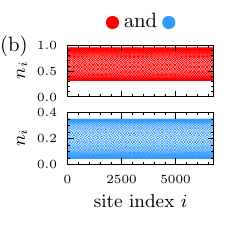} &
      \includegraphics[width=0.49\linewidth]{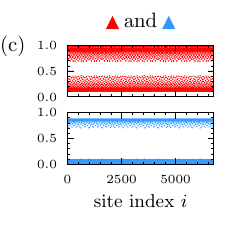} \\[-5pt]
      \includegraphics[width=0.49\linewidth]{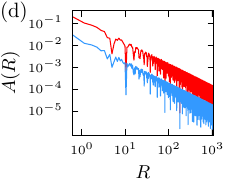} &
      \includegraphics[width=0.49\linewidth]{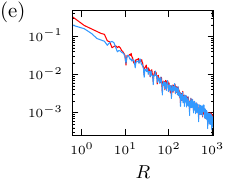} \\[0pt]
      \includegraphics[width=0.49\linewidth]{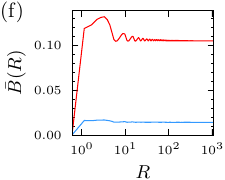} &
      \includegraphics[width=0.49\linewidth]{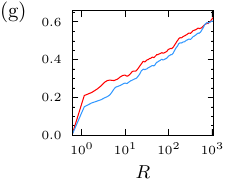}
    \end{tabular}
  \end{minipage}
  \caption{\label{fig:figure_A2}\protect\raggedright
    (a) Fractal dimension $D_f$ landscape for $\lambda = 1.5$. Gray dashed curves denote the mobility edges.
    Four representative $(\alpha,\nu)$ points in the gapless phases are marked by the symbols in the panel (a). For those points, (b,c) the charge-density distributions (d,e) log-log plot of $A(R)$ as a function of $R$ and (f,g) $\bar{B}(R)$ as a function of $R$. Panels (b,d,f) represent cases where the states around the Fermi energy are extended, while (c,e,g) represent cases where those states are localized.
  }
\end{figure*}

\begin{figure}
  \includegraphics[width=0.99\columnwidth,trim=4pt 0 0 3pt,clip]{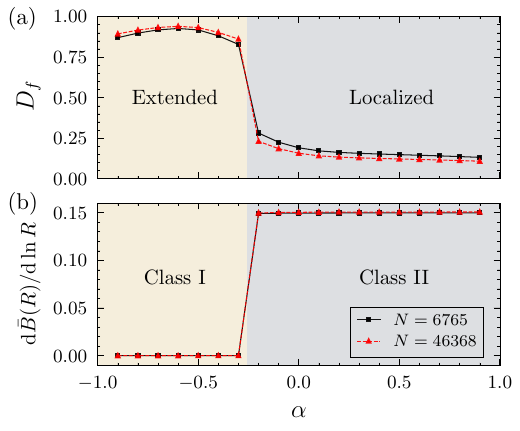}
  \begin{minipage}[b]{0.5\columnwidth}
    \centering
    \includegraphics[width=\linewidth]{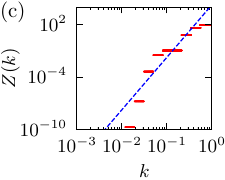}
  \end{minipage}\hfill
  \begin{minipage}[b]{0.5\columnwidth}
    \centering
    \includegraphics[width=\linewidth]{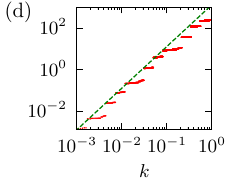}
  \end{minipage}
  \caption{\label{fig:figure_C1}\protect\raggedright Extended-to-localized transition accompanied by a change in the HU class, calculated for $\lambda=2.5$ and $\nu=2/3$. (a) Fractal dimension $D_f$ of the highest occupied state and (b) $\mathrm{d}\bar{B}(R)/\mathrm{d}\ln R$ obtained from $R\in(900,1000)$ as a function of $\alpha$. (c,d) Integrated intensity function $Z(k)$ for $\alpha=-0.3$ and $\alpha=-0.2$, respectively, calculated with $N=46368$ sites. The blue dashed line in (c) represents $Z(k)\sim k^{6}$ and the green dashed line in (d) represents $Z(k)\sim k^{2}$.
  }
\end{figure}

\subsection{Spectrum and fractal dimension}
\label{subsec:results_Df}

We begin by reviewing the landscape of the spectrum and the nature of the single-particle eigenstates in the GAA model. Unlike the Aubry--Andr\'e--Harper model ($\alpha=0$), where all the eigenstates are extended (localized) for $\lambda<2$ ($>2$), the GAA model at $\alpha\neq 0$ can have both extended and localized states at the same $\lambda$, with the mobility edge between them~\cite{PhysRevLett.114.146601}. As an example, we present the spectrum and the fractal dimension of the eigenstates for $\lambda=2.5$ [see Fig.~\ref{fig:figure_A1}(a)]. The eigenstates are localized (extended) in the black (gold) regions, as demonstrated by the value of the fractal dimension $D_f\simeq 0$ ($1$). The mobility edge given by Eq.~(\ref{eq:mobility_edge}) forms a boundary that separates the extended and localized regimes.

Figures~\ref{fig:figure_A1}(b--d) show the highest occupied single-particle states for fixed filling fraction $\nu$ and $\lambda$, illustrating three distinct localization regimes—extended, critical, and localized, respectively—as $\alpha$ is varied.
These results suggest that by tuning $\alpha$, which controls the relative position of the mobility edge with respect to the Fermi energy, one can place states with different localization characteristics in the vicinity of the Fermi energy. Note that a critical state emerges in the vicinity of the mobility edge [see the red star in Fig.~\ref{fig:figure_A1}(a)].

\begin{figure}
  \centering
  \begin{minipage}[b]{0.5\columnwidth}
    \centering
    \includegraphics[width=\linewidth,trim=0 0 0 5pt,clip]{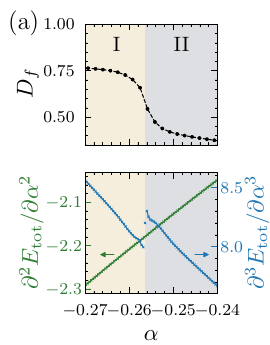}
  \end{minipage}\hfill
  \begin{minipage}[b]{0.5\columnwidth}
    \centering
    \includegraphics[width=\linewidth,trim=0 0 0 5pt,clip]{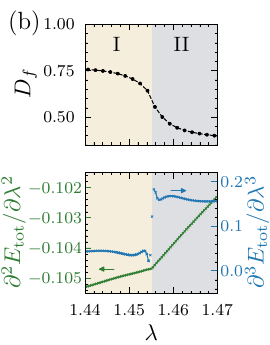}
  \end{minipage}
  \caption{\label{fig:figure_C2}\raggedright
  Fractal dimension $D_f$ and partial derivatives of the total energy per site $E_{\mathrm{tot}}$ across the transition between class-I and class-II HU at filling fraction $\nu = 2/3$. (a)~Dependence on $\alpha$ at fixed $\lambda=2.5$. (b) Dependence on $\lambda$ at fixed $\alpha=0.3$. The top panels show $D_f$ of the highest occupied state, while the bottom panels show the second and third derivatives of $E_{\mathrm{tot}}$ with respect to the tuning parameter. The yellow (gray) background color indicates the class-I (-II) HU.}
\end{figure}

\subsection{Hyperuniform charge distributions}
\label{subsec:results_HU}

Guided by the single-particle localization characteristics studied above,
we next investigate the many-body charge distribution $n_i$ and its HU.
Representative $(\alpha,\nu)$ points are selected from both gapped and gapless regions, as marked by different symbols in Fig.~\ref{fig:figure_A1}(a). Note that these points are selected so as to cover distinct combinations of localization characteristics of occupied states, as elaborated below.

Figures~\ref{fig:figure_B1}(a--d) display the charge distributions of different filling fractions and $\alpha$ values for $\lambda=2.5$. First of all, we find that all these charge distributions are hyperuniform, as $A(R)$ vanishes as $R$ increases [see Figs.~\ref{fig:figure_B1}(e--h)].
These results indicate that the charge distributions in the GAA model are generally hyperuniform, regardless of the presence of a gap or the localization properties of the occupied states.
Next, we shall examine the HU class of these distributions.
Note that the charge distributions for $\lambda<2$ ($\lambda\geq 2$) are always class-I (-II) hyperuniform for $\alpha=0$ in the gapless phases~\cite{PhysRevResearch.4.033241}.
To see whether the localization properties of the occupied states or the value of $\lambda$ itself determines the HU class of charge distribution, we consider two different $\lambda$ values, $\lambda=2.5$ and $\lambda=1.5$, in the following.

Figures~\ref{fig:figure_B1}(i--l) show $\bar{B}(R)$ as a function of $R$ for different $(\alpha,\nu)$ points for $\lambda=2.5$.
In the gapped phases, the charge distribution [Figs.~\ref{fig:figure_B1}(a,b)] always exhibits class-I HU, regardless of the localization properties of the highest occupied state and of the lowest unoccupied state.
We demonstrate this by studying four possible combinations of the localization properties of the highest occupied state and of the lowest unoccupied state: both localized (purple square), both extended (magenta square), localized--extended (green diamond), and extended--localized (brown diamond).
In all these four cases, the local charge density $n_i$ takes almost all values between its minimum to the maximum without a finite jump [compare Figs.~\ref{fig:figure_B1}(a,b) to \ref{fig:figure_B1}(d)].
The absence of the jump leads to small density fluctuations,  resulting in a class-I HU, as $\bar B(R)$ approaches a constant value with increasing $R$ [see Figs.~\ref{fig:figure_B1}(i,j)].

In the gapless phases, however, the HU class is influenced by the localization properties of occupied states.
Surprisingly, the HU class of gapless phase is determined solely by the localization properties of the states around the Fermi energy, even though the charge distribution contains contribution from all single-particle states below the Fermi energy.
Specifically, when the states around the Fermi energy are extended, the charge distribution [Figs.~\ref{fig:figure_B1}(c,d)] exhibits class-I HU, whereas when these states are localized, it exhibits class-II HU.
We examine this by considering four possibilities of the occupied states: (i) all extended (blue circle), (ii) all localized (blue triangle), (iii) the states around the Fermi energy are extended while localized states also exist below the Fermi energy (red circle), and (iv) the states around the Fermi energy are localized while extended states also exist below the Fermi energy (red triangle).
The latter two cases, (iii) and (iv), correspond to situations with a mobility edge below the Fermi energy, which is absent in the Aubry--Andr\'e--Harper model.
For cases (i) and (iii), $n_i$ takes almost all values between its minimum, whereas for cases (ii) and (iv), it exhibits a jump [compare Figs.~\ref{fig:figure_B1}(c) and \ref{fig:figure_B1}(d)].
This difference originates from the localization properties of the states around the Fermi energy and leaves a signature in the behavior of $\bar B(R)$. In detail, $\bar B(R)$ approaches a finite value (i.e., class-I HU) in the cases (i) and (iii),
whereas it grows logarithmically at large $R$ (i.e., class-II HU) in the cases (ii) and (iv)
[compare Figs.~\ref{fig:figure_B1}(k) and \ref{fig:figure_B1}(l)].

Next we study a subtle case that critical states appear around the Fermi energy. We investigate the situation where the mobility edge lies between the highest occupied state and the lowest unoccupied state, which is indicated by the red star in Fig.~\ref{fig:figure_A1}(a). In this case, the charge distribution covers almost all values from its minimum to maximum, while the number of sites with $n_i \approx 0.55$ is suppressed
[compare Fig.~\ref{fig:figure_critical}(a) and Figs.~\ref{fig:figure_B1}(a,d)]. This
suppression of the intermediate values of $n_i$ enhances density fluctuations, leading to class-II HU, as characterized by the logarithmic growth of $\bar{B}(R)$ for large $R$ [see Fig.~\ref{fig:figure_critical}(b)].

We also examine the case for $\lambda=1.5$, to ensure that the HU class is determined by the localization property of the states rather
than by the $\lambda$ value itself.
We draw the landscape of $D_f$ and HU analysis of several representative charge distributions in Fig.~\ref{fig:figure_A2}(a) and Figs.~\ref{fig:figure_A2}(b--g), respectively. In the gapped phases, the charge distribution shows class-I HU in all cases (not shown).
In the gapless phases, the HU class of the charge distribution [Figs.~\ref{fig:figure_A2}(b,c)] is determined solely by the states around the Fermi energy, as extended (localized) states correspond to class-I (-II) HU [see Figs.~\ref{fig:figure_A2}(d--g)].

\subsection{Phase transition}
\label{subsec:results_phase_transition}

As shown in the previous section, in gapless phases, class-I (-II) HU appear when the states around the Fermi energy are extended (localized).
Therefore, in gapless phases, the HU classification can be used as a diagnostic of the localization properties of these states.
Namely, without directly calculating the localization property, such as the localization length and conductance, one may be able to know it from the HU class of the charge distribution.
Figures~\ref{fig:figure_C1}(a) and \ref{fig:figure_C1}(b) show the fractal dimension of the highest occupied state, $D_f$, and $\mathrm{d}\bar{B}(R)/\mathrm{d}\ln R$ as a function of $\alpha$ for fixed $\lambda = 2.5$ and $\nu = 2/3$. As $\alpha$ increases, $D_f$ decreases abruptly, indicating a phase transition from an extended state to a localized state. Meanwhile, $\mathrm{d}\bar{B(R)}/\mathrm{d}\ln R$ introduced in Sec.~\ref{subsec:Model_and_Methods_HU} changes from zero to a finite value.
This indicates a transition from class-I to class-II HU (refer to Table~\ref{tab:hu_class}). Notably, the abrupt change in $D_f$ of the highest occupied state coincides with a transition in the HU class.
Hence, the HU class of charge distribution in the gapless phases serves as a probe of the localization properties around Fermi energy.

In Figs.~\ref{fig:figure_C1}(a) and \ref{fig:figure_C1}(b), we confirm the above results for a larger system size.
We compare the results of $D_f$ and $\mathrm{d}\bar{B}/\mathrm{d}\ln R$ between $N=F_{20}=6765$ and $N=F_{24}=46368$.
As the system size increases, $D_f$ approaches unity (zero), indicating the extended (localized) states in the thermodynamic limit.
Meanwhile, $\mathrm{d}\bar{B}/\mathrm{d}\ln R$, which captures the HU class, remains almost unchanged when the system size is increased.
This insensitivity to the system size suggests that the HU class can serve as a useful criterion for identifying the localization properties of the states around the Fermi energy.

We also examine the HU classes through the integrated intensity function $Z(k)$ defined in Eq.~\eqref{eq:Z(k)}. Given $\lambda=2.5$ and $\nu=2/3$,
Figs.~\ref{fig:figure_C1}(c) and \ref{fig:figure_C1}(d) show the small-$k$ scaling behaviors of $Z(k)\sim k^\delta$ for $\alpha=-0.3$ and $-0.2$, respectively. The scaling exponent $\delta>2$ for $\alpha=-0.3$, while $\delta=2$ for $\alpha=-0.2$. According to Table~\ref{tab:hu_class}, these scaling exponents correspond to class-I and class-II HU, respectively. Note that this is consistent with the observation in Fig.~\ref{fig:figure_C1}(b).

We further investigate the nature and order of the phase transition between inhomogeneous charge distributions of different HU classes.
In the gapless phases, both the change of $\alpha$ and $\lambda$ could lead to phase transitions between class-I and class-II HU at a fixed filling fraction $\nu$. In both cases, not only $E_{\mathrm{tot}}$ but also its first and second derivatives remain continuous, whereas a jump appears in its third derivative [see Fig.~\ref{fig:figure_C2}]. This result indicates a third-order transition, similarly to that found at the self-dual point ($\lambda=2$) in the Aubry--Andr\'e model~\cite{PhysRevResearch.4.033241}.

%% file: sections/discussion.tex
\section{Discussion}
\label{sec:discussion}

\begin{figure}
  \centering
  \begin{minipage}[b]{0.495\columnwidth}
    \centering
    \includegraphics[width=\linewidth]{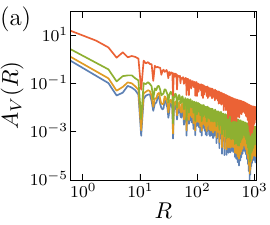}
  \end{minipage}\hfill
  \begin{minipage}[b]{0.495\columnwidth}
    \centering
    \includegraphics[width=\linewidth]{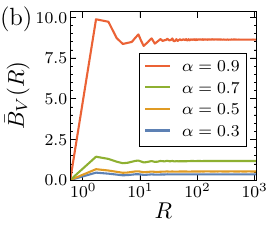}
  \end{minipage}
  \caption{\label{fig:figure_D1}\raggedright HU analysis of the on-site potential of the GAA model for $\lambda=2.5$ and different values of $\alpha$ [Eq.~\eqref{eq:gaa_potential}]. (a) $A_V(R)$ and (b) $\bar{B}_V(R)$ as functions of $R$.}
\end{figure}

While the charge distribution under a random potential is non-hyperuniform \cite{PhysRevResearch.4.033241}, that under the GAA potential is hyperuniform. The latter will be attributed to the HU of the GAA potential, as shown in Fig.~\ref{fig:figure_D1}(a), where $A_V(R)=\sigma_V^2(R)/R$ with $\sigma_V^2(R)=\overline{N_V(R)^2}-{\overline{N_V(R)}}^2$ and $N_V(R,r_c)=\sum_i V_i\,\Theta\!\left(R-|r_i-r_c|\right)$ goes to zero as $R$ increases. Figure \ref{fig:figure_D1}(b) shows that its HU class is I, as $\bar{B}_V(R)=\frac{1}{R} \int_0^R \sigma_V^2(R') \, dR'$ converges to a constant as $R$ increases. It is interesting that the charge distribution can show both classes I and II, depending on the model parameters, even though the potential is always class-I HU.


In detail, charge distributions always show
the class-I HU in all gapped phases, while show either class I or II in the gapless phases. Interestingly, as long as the charge distribution is contributed from
all the states within a ``band'' separated by finite gaps in the thermodynamic limit, it shows the class-I HU, irrespective of the localization properties of the occupied states~\cite{Jeon2026}.
On the other hand, the charge distribution is contributed from a partially filled band in the gapless phases. In this case, the localization properties of low-energy single-particle states influence the HU class, where extended (localized) states give class-I (-II) HU.

A particularly intriguing implication of our results is that the HU framework distinguishes insulating phases according to their microscopic origin. Notably, gapped phases are universally classified as class-I HU, whereas gapless localized phases belong to class-II HU, despite both being insulating from the transport perspective. This observation indicates that HU classification is sensitive not merely to the absence of conductivity, but also to the underlying mechanism responsible for insulation: whether it originates from a spectral gap, characterized by the absence of states at the Fermi level, or from a mobility gap induced by wave localization. In this sense, the long-range scaling of charge fluctuations provides a unified framework that encodes information conventionally separated into the density of states and transport properties. This interpretation is further consistent with previous studies~\cite{PhysRevResearch.4.033241} showing that Anderson insulators exhibit non-hyperuniform charge distributions, suggesting that long-range charge fluctuations are progressively more suppressed as one moves from mobility-gap-driven to spectral-gap-driven insulation. Moreover, the degree of suppression is expected to reflect the HU of the underlying potential landscape (refer to \cite{PhysRevResearch.4.033241} and see Fig.~\ref{fig:figure_D1}). Thus, the HU framework essentially provides a unified perspective that distinguishes not only whether a system is insulating, but also why it is insulating. Note that the HU framework can distinguish insulating phases of different origins because localized states behave as classical particles that have lost their wave character. By contrast, because the present framework neglects quantum fluctuations, it cannot distinguish gapped and gapless phases within extended regimes, where wave properties dominate. Hence, extending the HU framework to include quantum fluctuations and applying it to the GAA model would provide a route toward a finer classification of delocalized quantum phases, as suggested in Ref.~\cite{arxiv:2601.18331}.

On the phase transition between different HU classes, our results show that it is of the third order in the gapless region.
This may be compared with the transition between gapped and gapless phases in the Aubry--Andr\'e model, which is of the first order~\cite{PhysRevResearch.4.033241}.
In the latter case, the states at the Fermi energy disappear abruptly when the Fermi energy enters a gap, resulting in a kink in the $E_F$ dependence of $E_{\mathrm{tot}}$ and a jump in its first derivative. In contrast, the extended--localized transition within the gapless phase occurs when the Fermi energy coincides with the mobility edge, where the corresponding state is critical. Unlike the disappearance of states at the Fermi energy in a gapped phase, this critical state still contributes to the total energy, leading to a more smooth $E_{\mathrm{tot}}$.
Moreover, the singular-continuous density of states, associated with the critical state, would suppress fluctuations compared to a continuous density of states, which may result in a third-order transition.

%% file: sections/summary_and_outlook.tex
\section{Summary and Outlook}
\label{sec:summary_and_outlook}

We have applied the hyperuniformity framework to the charge distribution of the generalized Aubry--Andr\'e model, where extended and localized single-particle states can coexist below the Fermi energy.
Our results show that the charge distribution is always hyperunifom while its hyperuniformity class changes with the model parameters. In the gapped phases, the charge distribution always exhibits class-I hyperuniformity.
On the other hand, in the gapless phases, the localization properties of the states around the Fermi energy determine the hyperuniformity class: extended (localized) states give class-I (-II) hyperuniformity. Thus, the hyperuniformity class can be regarded as a direct fingerprint of the localization properties of states around the Fermi energy in the gapless phases. We further studied the phase transition when hyperuniformity class changes within the gapless phase and found it to be of the third order.

Our results suggest an intimate relation between many-body inhomogeneous charge distributions and localization properties of single-particle states around the Fermi level, which opens several promising avenues for future research. One important direction is to apply the quantum hyperuniformity framework~\cite{arxiv:2601.18331} to the generalized Aubry--Andr\'e model, which would help establish a more comprehensive hyperuniformity classification, as discussed in Sec.~\ref{sec:discussion}.
Another interesting direction is to apply the present hyperuniformity analysis to quasiperiodic models in higher dimensions~\cite{PhysRevB.101.014205,PhysRevB.109.014210,PhysRevB.96.214201,PhysRevB.106.104205} and with interactions~\cite{JPSJ.84.023701,PhysRevB.102.195142,PhysRevB.96.214402,PhysRevB.105.205138,JPSJ.59.811,JPSJ.93.114005,PhysRevLett.86.1331,PhysRevA.82.043613}.
These systems would provide useful platforms for testing the generality of our conclusion. Beyond quasiperiodic systems, disordered hyperuniform systems can also induce delocalized single-particle states and mobility edges~\cite{PhysRevB.113.L020201,PhysRevB112.L161104,MaterTrans67.680}, and hence offer another intriguing subject of the hyperuniformity analysis.
Phase transitions and associated critical behaviors in these models also represent interesting future research topics.
Finally, beyond charge distributions, the hyperuniformity framework may also be applied to other local observables, such as superconducting order parameters \cite{PhysRevB.95.024509,PhysRevB.100.014510,PhysRevResearch.1.022002,PhysRevB.112.174511} and local magnetic moment \cite{PhysRevLett.90.177205,PhysRevB.75.212407,10.1140/epjb/e2012-21003-x,PhysRevB.96.214402,PhysRevB.102.115125}. Such studies may provide further insight into possible relations between many-body spatial patterns and underlying single-particle properties.

%% file: sections/appendix.tex
\section{Model with multiple mobility edges}
\label{sec:Appendix}

\begin{figure}
  \includegraphics[width=0.99\columnwidth,trim=0 0 0 3pt,clip]{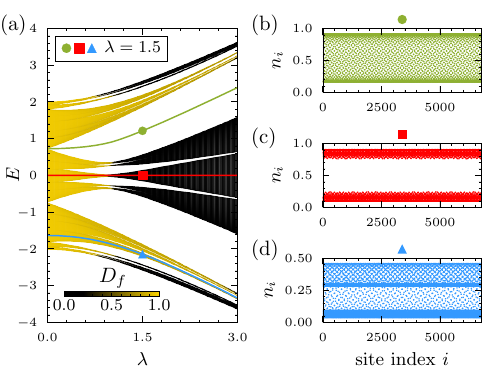}\\
  \vspace{-3pt}
  \includegraphics[width=0.99\columnwidth,trim=2pt 0 2pt 2pt,clip]{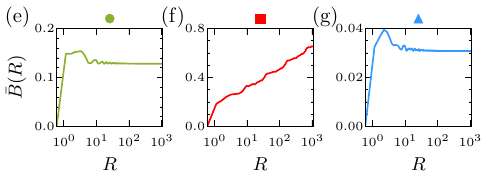}
  \caption{\label{fig:figure_Appendix_1}\protect\raggedright (a) Fractal dimension $D_f$ landscape for the model described in Appendix \ref{sec:Appendix} with $\eta=3$. From top to bottom in panel (a), the green, red, and blue curves correspond to different filling fractions: $\nu=F_{19}/F_{20}$, $\nu=\frac{N-1}{2N}$, and $\nu=8/41$, respectively. Different symbols indicate representative points at $\lambda=1.5$ for these filling fractions in regions with distinct characteristics. For these points, (b--d) charge-density distributions and (e--g) $\bar B(R)$ as a function of $R$. $N=F_{20}=6765$ is used.}
\end{figure}

\begin{figure}
  \includegraphics[width=0.99\columnwidth]{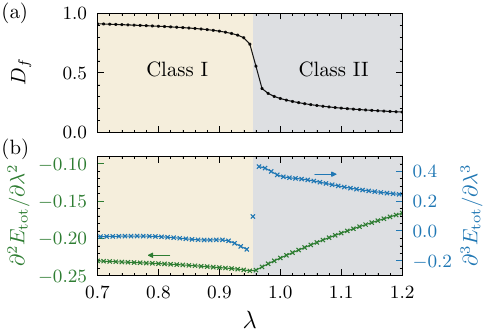}\\
  \caption{\label{fig:figure_Appendix_2}\protect\raggedright (a) Fractal dimension $D_f$ and (b) partial derivatives of the total energy per site $E_{\mathrm{tot}}$ for $\nu=\frac{N-1}{2N}$ and $\eta=3$, along the red line in Fig.~\ref{fig:figure_Appendix_1}. The yellow (gray) background color indicates the class-I (-II) HU. $N=6765$ is used.}
\end{figure}

Since the GAA model exhibits at most
one mobility edge for each $(\lambda,\alpha)$, here we examine the validity of our results in a model that exhibits multiple mobility edges. We consider the model proposed by Hiramoto and Kohmoto in Ref.~\cite{PhysRevLett.62.2714}.
This model follows the same Hamiltonian in Eq.~\eqref{eq:gaa_H} but with the potential given by~\cite{PhysRevLett.62.2714}
\begin{equation}
  V_i^{\mathrm{HK}}=\lambda\,\frac{\tanh\left[\eta\cos\left(2\pi \beta i\right)\right]}{\tanh{\eta}}.
  \label{eq:tanh_potential}
\end{equation}
This model interpolates between the Aubry--Andr\'e model and the diagonal Fibonacci model as the parameter $\eta>0$ varies: when $\eta$ approaches $0$, it is reduced to the Aubry--Andr\'e model, whereas when $\eta$ goes to infinity, it approaches the diagonal Fibonacci model~\cite{PhysRevLett.62.2714,PhysRevLett.50.1870,RevModPhys.93.045001}.

We draw the fractal dimension landscape,
charge distributions, and $\bar B(R)$ for $\eta =3$ in Figs.~\ref{fig:figure_Appendix_1}(a), \ref{fig:figure_Appendix_1}(b--d), and \ref{fig:figure_Appendix_1}(e--g), respectively. In Fig.~\ref{fig:figure_Appendix_1}(a), the eigenstates are localized (extended) in the black (gold) regions, corresponding to $D_f \simeq 0$ ($1$).
Notably, Fig.~\ref{fig:figure_Appendix_1}(a) shows that for $\lambda\gtrsim1.5$ the localization characteristics (measured by $D_f$) alternate as localized $\to$ extended $\to$ localized $\to$ extended $\to$ localized from bottom to top energies. This implies the existence of at least four mobility edges in the spectrum between localized and extended regions (i.e., multiple mobility edges).
For $\lambda=1.5$, we choose three representative filling fractions $\nu$ with different characteristics: gapped phase (green circle), gapless phase with localized low-energy states (red square), gapless phase with extended low-energy states (blue triangle).
The corresponding charge distributions [Figs.~\ref{fig:figure_Appendix_1}(b--d)] show a jump only for the red square and no jump for the other two cases.
We find that all these charge distributions show HU, which may be expected from the fact that the charge distributions are hyperuniform in both the Aubry--Andr\'e and the Fibonacci models~\cite{PhysRevResearch.4.033241}.
In detail, class I appears in the gapped phases [see Figs.~\ref{fig:figure_Appendix_1}(b,e)], whereas different classes appear in the gapless phases. The class in the gapless phases is determined solely by the low-energy states: extended (localized) states correspond to class-I (-II) HU [see Figs.~\ref{fig:figure_Appendix_1}(f,g)].

Next, we examine the order of the extended-to-localized phase transition in the gapless phases for a fixed $\nu = \frac{N-1}{2N}$ with varying $\lambda$ [along the red line in Fig.~\ref{fig:figure_Appendix_1}(a)].
Figure~\ref{fig:figure_Appendix_2}(a) shows that $D_f$ steeply changes around $\lambda=0.96$. Correspondingly, the third derivative of the total energy $E_{\rm tot}$ with respect to $\lambda$ jumps, as shown in Fig.~\ref{fig:figure_Appendix_2}(b). Namely, the transition is of the third order.
These observations corroborate our conclusions in the main text.

%% file: refs.bib
@article{PhysRevLett.114.146601,
  title = {Nearest Neighbor Tight Binding Models with an Exact Mobility Edge in One Dimension},
  author = {Ganeshan, Sriram and Pixley, J. H. and Das Sarma, S.},
  journal = {Phys. Rev. Lett.},
  volume = {114},
  issue = {14},
  pages = {146601},
  numpages = {5},
  year = {2015},
  month = {Apr},
  publisher = {American Physical Society},
  doi = {10.1103/PhysRevLett.114.146601},
  url = {https://link.aps.org/doi/10.1103/PhysRevLett.114.146601}
}

@article{PhysRevResearch.4.033241,
  title = {Quantum phase transition between hyperuniform density distributions},
  author = {Sakai, Shiro and Arita, Ryotaro and Ohtsuki, Tomi},
  journal = {Phys. Rev. Res.},
  volume = {4},
  issue = {3},
  pages = {033241},
  numpages = {15},
  year = {2022},
  month = {Sep},
  publisher = {American Physical Society},
  doi = {10.1103/PhysRevResearch.4.033241},
  url = {https://link.aps.org/doi/10.1103/PhysRevResearch.4.033241}
}

@article{arxiv:2601.18331,
  title = {Quantum Hyperuniformity and Quantum Weight}, 
  author = {Junmo Jeon and Shiro Sakai},
  journal = {arXiv:2601.18331},
  year = {2026},
  url = {https://arxiv.org/abs/2601.18331}, 
}

@article{Aubry1980,
  author = {Aubry, Serge and André, Gilles},
  year = {1980},
  month = {01},
  pages = {},
  title = {Analyticity breaking and Anderson localization in incommensurate lattices},
  volume = {3},
  journal = {Proceedings, VIII International Colloquium on Group-Theoretical Methods in Physics}
}

@article{PhysRevE.68.041113,
  title = {Local density fluctuations, hyperuniformity, and order metrics},
  author = {Torquato, Salvatore and Stillinger, Frank H.},
  journal = {Phys. Rev. E},
  volume = {68},
  issue = {4},
  pages = {041113},
  numpages = {25},
  year = {2003},
  month = {Oct},
  publisher = {American Physical Society},
  doi = {10.1103/PhysRevE.68.041113},
  url = {https://link.aps.org/doi/10.1103/PhysRevE.68.041113}
}

@article{TORQUATO20181,
  title = {Hyperuniform states of matter},
  journal = {Physics Reports},
  volume = {745},
  pages = {1-95},
  year = {2018},
  issn = {0370-1573},
  doi = {https://doi.org/10.1016/j.physrep.2018.03.001},
  url = {https://www.sciencedirect.com/science/article/pii/S037015731830036X},
  author = {Salvatore Torquato}
}

@article{Torquato2026Weighted,
  title = {Hyperuniformity of Weighted Particle Systems},
  author = {Torquato, Salvatore and Kim, Jaeuk and Klatt, Michael A. and Car, Roberto and Steinhardt, Paul J.},
  journal = {Physical Review X},
  volume = {16},
  number = {1},
  pages = {011042},
  year = {2026},
  publisher = {American Physical Society},
  doi = {10.1103/fr99-qh7h},
  url = {https://doi.org/10.1103/fr99-qh7h}
}

@article{harper-1955,
	author = {Harper, P G},
	journal = {Proceedings of the Physical Society Section A},
	month = {10},
	number = {10},
	pages = {874--878},
	title = {{Single band motion of conduction electrons in a uniform magnetic field}},
	volume = {68},
	year = {1955},
	doi = {10.1088/0370-1298/68/10/304},
	url = {https://doi.org/10.1088/0370-1298/68/10/304},
}

@article{Kamiya2018,
  title = {Discovery of superconductivity in quasicrystal},
  author = {Kamiya, K. and Takeuchi, T. and Kabeya, N. and Wada, N. and Ishimasa, T. and Ochiai, A. and Deguchi, K. and Imura, K. and Sato, N. K.},
  journal = {Nat. Commun.},
  volume = {9},
  pages = {154},
  year = {2018},
  doi = {10.1038/s41467-017-02667-x},
  url = {https://doi.org/10.1038/s41467-017-02667-x}
}

@article{Tokumoto2024,
  title = {Superconductivity in a van der {Waals} layered quasicrystal},
  author = {Tokumoto, Yuki and Hamano, Kotaro and Nakagawa, Sunao and Kamimura, Yasushi and Suzuki, Shintaro and Tamura, Ryuji and Edagawa, Keiichi},
  journal = {Nat. Commun.},
  volume = {15},
  pages = {1529},
  year = {2024},
  doi = {10.1038/s41467-024-45952-2},
  url = {https://doi.org/10.1038/s41467-024-45952-2}
}

@article{Tamura2021,
  title = {Experimental Observation of Long-Range Magnetic Order in Icosahedral Quasicrystals},
  author = {Tamura, Ryuji and Ishikawa, Asuka and Suzuki, Shintaro and Kotajima, Takahiro and Tanaka, Yuki and Seki, Toshiki and Shibata, Kenya and Yamada, Tomohiro and Fujii, Takenori and Wang, Chin-Wei and Avdeev, Maxim and Nawa, Kazuhiro and Okuyama, Daisuke and Sato, Taku J.},
  journal = {J. Am. Chem. Soc.},
  volume = {143},
  number = {47},
  pages = {19938--19944},
  year = {2021},
  doi = {10.1021/jacs.1c09954},
  url = {https://doi.org/10.1021/jacs.1c09954}
}

@article{Tamura2025,
  title = {Observation of antiferromagnetic order in a quasicrystal},
  author = {Tamura, R. and Abe, T. and Yoshida, S. and Shimozaki, Y. and Suzuki, S. and Ishikawa, A. and Labib, F. and Avdeev, M. and Kinjo, K. and Nawa, K. and Sato, T. J.},
  journal = {Nat. Phys.},
  volume = {21},
  number = {6},
  pages = {974--979},
  year = {2025},
  doi = {10.1038/s41567-025-02858-0},
  url = {https://doi.org/10.1038/s41567-025-02858-0}
}

@article{Landau1937,
  title = {On the theory of phase transitions},
  author = {Landau, L. D.},
  journal = {Zh. Eksp. Teor. Fiz.},
  volume = {7},
  pages = {19--32},
  year = {1937}
}

@article{PhysRevB.105.054202,
  title = {Hyperuniform electron distributions controlled by electron interactions in quasicrystals},
  author = {Sakai, Shiro and Arita, Ryotaro and Ohtsuki, Tomi},
  journal = {Phys. Rev. B},
  volume = {105},
  pages = {054202},
  year = {2022},
  doi = {10.1103/PhysRevB.105.054202},
  url = {https://doi.org/10.1103/PhysRevB.105.054202}
}

@article{PhysRevB.105.205138,
  title = {Doped Mott insulator on a Penrose tiling},
  author = {Sakai, Shiro and Takemori, Nayuta},
  journal = {Phys. Rev. B},
  volume = {105},
  issue = {20},
  pages = {205138},
  year = {2022},
  month = {May},
  publisher = {American Physical Society},
  doi = {10.1103/PhysRevB.105.205138},
  url = {https://doi.org/10.1103/PhysRevB.105.205138}
}

@article{PhysRevB.95.024509,
  title = {Superconductivity on a quasiperiodic lattice: Extended-to-localized crossover of Cooper pairs},
  author = {Sakai, Shiro and Takemori, Nayuta and Koga, Akihisa and Arita, Ryotaro},
  journal = {Phys. Rev. B},
  volume = {95},
  issue = {2},
  pages = {024509},
  year = {2017},
  month = {Jan},
  publisher = {American Physical Society},
  doi = {10.1103/PhysRevB.95.024509},
  url = {https://doi.org/10.1103/PhysRevB.95.024509}
}

@article{PhysRevB.100.014510,
  title = {Conventional superconductivity in quasicrystals},
  author = {Ara'ujo, Ronaldo N. and Andrade, Eric C.},
  journal = {Phys. Rev. B},
  volume = {100},
  issue = {1},
  pages = {014510},
  year = {2019},
  month = {Jul},
  publisher = {American Physical Society},
  doi = {10.1103/PhysRevB.100.014510},
  url = {https://doi.org/10.1103/PhysRevB.100.014510}
}

@article{PhysRevResearch.1.022002,
  title = {Exotic pairing state in quasicrystalline superconductors under a magnetic field},
  author = {Sakai, Shiro and Arita, Ryotaro},
  journal = {Phys. Rev. Research},
  volume = {1},
  issue = {2},
  pages = {022002},
  year = {2019},
  month = {Sep},
  publisher = {American Physical Society},
  doi = {10.1103/PhysRevResearch.1.022002},
  url = {https://doi.org/10.1103/PhysRevResearch.1.022002}
}

@article{PhysRevB.112.174511,
  title = {Interaction-tuned pattern-selective superconductivity: Application to the dodecagonal quasicrystal},
  author = {Jeon, Junmo and Lee, SungBin},
  journal = {Phys. Rev. B},
  volume = {112},
  issue = {17},
  pages = {174511},
  year = {2025},
  month = {Nov},
  publisher = {American Physical Society},
  doi = {10.1103/PhysRevB.112.174511},
  url = {https://doi.org/10.1103/PhysRevB.112.174511}
}

@article{PhysRevLett.90.177205,
  title = {Quantum Antiferromagnetism in Quasicrystals},
  author = {Wessel, Stefan and Jagannathan, Anuradha and Haas, Stephan},
  journal = {Phys. Rev. Lett.},
  volume = {90},
  issue = {17},
  pages = {177205},
  year = {2003},
  month = {May},
  publisher = {American Physical Society},
  doi = {10.1103/PhysRevLett.90.177205},
  url = {https://doi.org/10.1103/PhysRevLett.90.177205}
}

@article{PhysRevB.75.212407,
  title = {Penrose quantum antiferromagnet},
  author = {Jagannathan, Anuradha and Szallas, Attila and Wessel, Stefan and Duneau, Michel},
  journal = {Phys. Rev. B},
  volume = {75},
  issue = {21},
  pages = {212407},
  year = {2007},
  month = {Jun},
  publisher = {American Physical Society},
  doi = {10.1103/PhysRevB.75.212407},
  url = {https://doi.org/10.1103/PhysRevB.75.212407}
}

@article{10.1140/epjb/e2012-21003-x,
  title = {Quasiperiodic Heisenberg antiferromagnets in two dimensions},
  author = {Jagannathan, A.},
  journal = {Eur. Phys. J. B},
  volume = {85},
  number = {2},
  pages = {68},
  year = {2012},
  doi = {10.1140/epjb/e2012-21003-x},
  url = {https://doi.org/10.1140/epjb/e2012-21003-x}
}

@article{PhysRevB.96.214402,
  title = {Antiferromagnetic order in the Hubbard model on the Penrose lattice},
  author = {Koga, Akihisa and Tsunetsugu, Hirokazu},
  journal = {Phys. Rev. B},
  volume = {96},
  issue = {21},
  pages = {214402},
  year = {2017},
  month = {Dec},
  publisher = {American Physical Society},
  doi = {10.1103/PhysRevB.96.214402},
  url = {https://doi.org/10.1103/PhysRevB.96.214402}
}

@article{PhysRevB.102.115125,
  title = {Superlattice structure in the antiferromagnetically ordered state in the Hubbard model on the Ammann--Beenker tiling},
  author = {Koga, Akihisa},
  journal = {Phys. Rev. B},
  volume = {102},
  issue = {11},
  pages = {115125},
  year = {2020},
  month = {Sep},
  publisher = {American Physical Society},
  doi = {10.1103/PhysRevB.102.115125},
  url = {https://doi.org/10.1103/PhysRevB.102.115125}
}

@article{PhysRevB.95.054119,
  title = {Hyperuniformity of quasicrystals},
  author = {O\ifmmode \breve{g}\else \u{g}\fi{}uz, Erdal C. and Socolar, Joshua E. S. and Steinhardt, Paul J. and Torquato, Salvatore},
  journal = {Phys. Rev. B},
  volume = {95},
  issue = {5},
  pages = {054119},
  numpages = {10},
  year = {2017},
  month = {Feb},
  publisher = {American Physical Society},
  doi = {10.1103/PhysRevB.95.054119},
  url = {https://link.aps.org/doi/10.1103/PhysRevB.95.054119}
}

@article{PhysRevLett.125.196604,
  title = {One-Dimensional Quasiperiodic Mosaic Lattice with Exact Mobility Edges},
  author = {Wang, Yucheng and Xia, Xu and Zhang, Long and Yao, Hepeng and Chen, Shu and You, Jiangong and Zhou, Qi and Liu, Xiong-Jun},
  journal = {Phys. Rev. Lett.},
  volume = {125},
  issue = {19},
  pages = {196604},
  numpages = {6},
  year = {2020},
  month = {Nov},
  publisher = {American Physical Society},
  doi = {10.1103/PhysRevLett.125.196604},
  url = {https://link.aps.org/doi/10.1103/PhysRevLett.125.196604}
}

@article{PhysRevB.113.L020201,
  title = {Delocalization induced by enhanced hyperuniformity in one-dimensional disordered systems},
  author = {Jeon, Junmo and Ikeda, Harukuni and Sakai, Shiro},
  journal = {Phys. Rev. B},
  volume = {113},
  issue = {2},
  pages = {L020201},
  numpages = {7},
  year = {2026},
  month = {Jan},
  publisher = {American Physical Society},
  doi = {10.1103/mmw9-lvzd},
  url = {https://link.aps.org/doi/10.1103/mmw9-lvzd}
}

@article{PhysRevB.101.014205,
  title = {Mixed spectra and partially extended states in a two-dimensional quasiperiodic model},
  author = {Szab\'o, Attila and Schneider, Ulrich},
  journal = {Phys. Rev. B},
  volume = {101},
  issue = {1},
  pages = {014205},
  numpages = {8},
  year = {2020},
  month = {Jan},
  publisher = {American Physical Society},
  doi = {10.1103/PhysRevB.101.014205},
  url = {https://link.aps.org/doi/10.1103/PhysRevB.101.014205}
}

@article{PhysRevB.109.014210,
  title = {Critical states and anomalous mobility edges in two-dimensional diagonal quasicrystals},
  author = {Duncan, Callum W.},
  journal = {Phys. Rev. B},
  volume = {109},
  issue = {1},
  pages = {014210},
  numpages = {11},
  year = {2024},
  month = {Jan},
  publisher = {American Physical Society},
  doi = {10.1103/PhysRevB.109.014210},
  url = {https://link.aps.org/doi/10.1103/PhysRevB.109.014210}
}

@article{PhysRevB.96.214201,
  title = {Anderson localization transitions with and without random potentials},
  author = {Devakul, Trithep and Huse, David A.},
  journal = {Phys. Rev. B},
  volume = {96},
  issue = {21},
  pages = {214201},
  numpages = {16},
  year = {2017},
  month = {Dec},
  publisher = {American Physical Society},
  doi = {10.1103/PhysRevB.96.214201},
  url = {https://link.aps.org/doi/10.1103/PhysRevB.96.214201}
}

@article{PhysRevB.102.195142,
  title = {Nonequilibrium steady state phases of the interacting Aubry--Andr\'e--Harper model},
  author = {Yoo, Yongchan and Lee, Junhyun and Swingle, Brian},
  journal = {Phys. Rev. B},
  volume = {102},
  issue = {19},
  pages = {195142},
  numpages = {10},
  year = {2020},
  month = {Nov},
  publisher = {American Physical Society},
  doi = {10.1103/PhysRevB.102.195142},
  url = {https://link.aps.org/doi/10.1103/PhysRevB.102.195142}
}

@article{Kosterlitz-Thouless1973,
  title = {Ordering, metastability and phase transitions in two-dimensional systems},
  author = {Kosterlitz, J. M. and Thouless, D. J.},
  journal = {J. Phys. C: Solid State Phys.},
  volume = {6},
  number = {7},
  pages = {1181--1203},
  year = {1973},
  doi = {10.1088/0022-3719/6/7/010},
  url = {https://doi.org/10.1088/0022-3719/6/7/010}
}

@article{Haldane1988,
  title = {Model for a Quantum Hall Effect without Landau Levels: Condensed-Matter Realization of the ``Parity Anomaly''},
  author = {Haldane, F. D. M.},
  journal = {Phys. Rev. Lett.},
  volume = {61},
  issue = {18},
  pages = {2015--2018},
  year = {1988},
  month = {Oct},
  publisher = {American Physical Society},
  doi = {10.1103/PhysRevLett.61.2015},
  url = {https://doi.org/10.1103/PhysRevLett.61.2015}
}

@article{Shechtmann1984,
  title = {Metallic Phase with Long-Range Orientational Order and No Translational Symmetry},
  author = {Shechtman, D. and Blech, I. and Gratias, D. and Cahn, J. W.},
  journal = {Phys. Rev. Lett.},
  volume = {53},
  issue = {20},
  pages = {1951--1953},
  year = {1984},
  month = {Nov},
  publisher = {American Physical Society},
  doi = {10.1103/PhysRevLett.53.1951},
  url = {https://doi.org/10.1103/PhysRevLett.53.1951}
}

@article{Levine-Steinhardt1984,
  title = {Quasicrystals: A New Class of Ordered Structures},
  author = {Levine, Dov and Steinhardt, Paul J.},
  journal = {Phys. Rev. Lett.},
  volume = {53},
  issue = {26},
  pages = {2477--2480},
  year = {1984},
  month = {Dec},
  publisher = {American Physical Society},
  doi = {10.1103/PhysRevLett.53.2477},
  url = {https://doi.org/10.1103/PhysRevLett.53.2477}
}

@article{PhysRevB.102.115108,
  title = {Physical properties of weak-coupling quasiperiodic superconductors},
  author = {Takemori, Nayuta and Arita, Ryotaro and Sakai, Shiro},
  journal = {Phys. Rev. B},
  volume = {102},
  issue = {11},
  pages = {115108},
  year = {2020},
  month = {Sep},
  publisher = {American Physical Society},
  doi = {10.1103/PhysRevB.102.115108},
  url = {https://doi.org/10.1103/PhysRevB.102.115108}
}

@article{PhysRevB.104.144511,
  title = {Topological superconductivity in quasicrystals},
  author = {Ghadimi, Rasoul and Sugimoto, Takanori and Tanaka, K. and Tohyama, Takami},
  journal = {Phys. Rev. B},
  volume = {104},
  issue = {14},
  pages = {144511},
  year = {2021},
  month = {Oct},
  publisher = {American Physical Society},
  doi = {10.1103/PhysRevB.104.144511},
  url = {https://doi.org/10.1103/PhysRevB.104.144511}
}

@article{PhysRevLett.80.2717,
  title = {Symmetry of Magnetically Ordered Quasicrystals},
  author = {Lifshitz, Ron},
  journal = {Phys. Rev. Lett.},
  volume = {80},
  issue = {12},
  pages = {2717--2720},
  year = {1998},
  month = {Mar},
  publisher = {American Physical Society},
  doi = {10.1103/PhysRevLett.80.2717},
  url = {https://doi.org/10.1103/PhysRevLett.80.2717}
}

@article{pnas.2112202118,
  title = {Topological magnetic textures and long-range orders in terbium-based quasicrystal and approximant},
  author = {Watanabe, Shinji},
  journal = {Proc. Natl. Acad. Sci. U.S.A.},
  volume = {118},
  number = {43},
  pages = {e2112202118},
  year = {2021},
  doi = {10.1073/pnas.2112202118},
  url = {https://doi.org/10.1073/pnas.2112202118}
}

@article{PhysRevMaterials.4.024417,
  title = {Magnetic orders induced by {RKKY} interaction in {Tsai}-type quasicrystalline approximant {Au-Al-Gd}},
  author = {Miyazaki, Haruka and Sugimoto, Takanori and Morita, Katsuhiro and Tohyama, Takami},
  journal = {Phys. Rev. Materials},
  volume = {4},
  issue = {2},
  pages = {024417},
  year = {2020},
  month = {Feb},
  publisher = {American Physical Society},
  doi = {10.1103/PhysRevMaterials.4.024417},
  url = {https://doi.org/10.1103/PhysRevMaterials.4.024417}
}

@article{PhysRevLett.62.2714,
  title = {New localization in a quasiperiodic system},
  author = {Hiramoto, Hisashi and Kohmoto, Mahito},
  journal = {Phys. Rev. Lett.},
  volume = {62},
  issue = {23},
  pages = {2714--2717},
  numpages = {0},
  year = {1989},
  month = {Jun},
  publisher = {American Physical Society},
  doi = {10.1103/PhysRevLett.62.2714},
  url = {https://link.aps.org/doi/10.1103/PhysRevLett.62.2714}
}

@article{PhysRevB.106.104205,
  title = {Universality classes of the Anderson transitions driven by quasiperiodic potential in the three-dimensional Wigner-Dyson symmetry classes},
  author = {Luo, Xunlong and Ohtsuki, Tomi},
  journal = {Phys. Rev. B},
  volume = {106},
  issue = {10},
  pages = {104205},
  numpages = {11},
  year = {2022},
  month = {Sep},
  publisher = {American Physical Society},
  doi = {10.1103/PhysRevB.106.104205},
  url = {https://link.aps.org/doi/10.1103/PhysRevB.106.104205}
}

@article{JPSJ.59.811,
  title = {The Effect of an Interelectron Interaction on Electronic Properties in One-Dimensional Quasiperiodic Systems},
  author = {Hiramoto, Hisashi},
  journal = {J. Phys. Soc. Jpn},
  volume = {59},
  number = {3},
  pages = {811--814},
  year = {1990},
  doi = {10.1143/JPSJ.59.811},
  url = {https://doi.org/10.1143/JPSJ.59.811}
}

@article{JPSJ.93.114005,
  author = {Hori ,Masahiro and Sugimoto ,Takanori and Hashizume ,Yoichiro and Tohyama ,Takami},
  title = {Multifractality and Hyperuniformity in Quasicrystalline Bose–Hubbard Models with and without Disorder},
  journal={J. Phys. Soc. Jpn},
  volume = {93},
  number = {11},
  pages = {114005},
  year = {2024},
  doi = {10.7566/JPSJ.93.114005},
  url = {https://doi.org/10.7566/JPSJ.93.114005},
}

@article{PhysRevLett.86.1331,
  title = {Quasiperiodic Hubbard Chains},
  author = {Hida, Kazuo},
  journal = {Phys. Rev. Lett.},
  volume = {86},
  issue = {7},
  pages = {1331--1334},
  numpages = {0},
  year = {2001},
  month = {Feb},
  publisher = {American Physical Society},
  doi = {10.1103/PhysRevLett.86.1331},
  url = {https://link.aps.org/doi/10.1103/PhysRevLett.86.1331}
}

@article{PhysRevA.82.043613,
  title = {Stability of the superfluid state in a disordered one-dimensional ultracold fermionic gas},
  author = {Tezuka, Masaki and Garc\'{\i}a-Garc\'{\i}a, Antonio M.},
  journal = {Phys. Rev. A},
  volume = {82},
  issue = {4},
  pages = {043613},
  numpages = {5},
  year = {2010},
  month = {Oct},
  publisher = {American Physical Society},
  doi = {10.1103/PhysRevA.82.043613},
  url = {https://link.aps.org/doi/10.1103/PhysRevA.82.043613}
}

@article{PhysRevB112.L161104,
  title = {Electron model on {Truchet} tiling: Extended-to-localized transitions and asymmetric spectrum},
  author = {Jeon, Junmo and Sakai, Shiro},
  journal = {Phys. Rev. B},
  volume = {112},
  issue = {16},
  pages = {L161104},
  numpages = {6},
  year = {2025},
  month = {Oct},
  publisher = {American Physical Society},
  doi = {10.1103/qg97-s1vt},
  url = {https://link.aps.org/doi/10.1103/qg97-s1vt}
}

@article{MaterTrans67.680,
  title = {Hyperuniform Structure and Mobility Edge in the Random {Truchet} Tiling},
  author = {Jeon, Junmo and Ikeda, Harukuni and Sakai, Shiro},
  journal = {Materials Transactions},
  volume = {67},
  issue = {5},
  pages = {680--685},
  year = {2026},
  month = {May},
  doi = {10.2320/matertrans.MT-MD2025006},
  url = {https://doi.org/10.2320/matertrans.MT-MD2025006}
}

@article{JPSJ.84.023701,
  title = {Local Electron Correlations in a Two-Dimensional {Hubbard} Model on the {Penrose} Lattice},
  author = {Takemori, Nayuta and Koga, Akihisa},
  journal = {J. Phys. Soc. Jpn},
  volume = {84},
  number = {2},
  pages = {023701},
  year = {2015},
  month = {Feb},
  doi = {10.7566/JPSJ.84.023701},
  url = {https://doi.org/10.7566/JPSJ.84.023701}
}

@article{RevModPhys.93.045001,
  title = {The Fibonacci quasicrystal: Case study of hidden dimensions and multifractality},
  author = {Jagannathan, Anuradha},
  journal = {Rev. Mod. Phys.},
  volume = {93},
  issue = {4},
  pages = {045001},
  numpages = {37},
  year = {2021},
  month = {Nov},
  publisher = {American Physical Society},
  doi = {10.1103/RevModPhys.93.045001},
  url = {https://link.aps.org/doi/10.1103/RevModPhys.93.045001}
}

@article{PhysRevLett.50.1870,
  title = {Localization Problem in One Dimension: Mapping and Escape},
  author = {Kohmoto, Mahito and Kadanoff, Leo P. and Tang, Chao},
  journal = {Phys. Rev. Lett.},
  volume = {50},
  issue = {23},
  pages = {1870--1872},
  numpages = {0},
  year = {1983},
  month = {Jun},
  publisher = {American Physical Society},
  doi = {10.1103/PhysRevLett.50.1870},
  url = {https://link.aps.org/doi/10.1103/PhysRevLett.50.1870}
}

@article{Jeon2026,
  journal = {{\it in preparation}},
  author = {
    Junmo Jeon and Shiro Sakai
  }
}
